\documentclass[prb,twocolumn,superscriptaddress,floatfix,graphicx,showpacs]{revtex4-1}
\usepackage{ifpdf}
 \newif\ifpdf
\ifx\pdfoutput\undefined
   \pdffalse
\else
   \pdfoutput=1
   \pdftrue
\fi
\ifpdf
   \usepackage{graphicx}
   \usepackage{epstopdf}
   \usepackage{color}
\else
   \usepackage{graphicx}
\fi
\usepackage{cancel}
\usepackage{srctex}
\usepackage{hyperref}

\usepackage{amsmath,amssymb}

\usepackage{epsfig}

\usepackage{epsfig}

\begin{document}

\title{Collective modes in simple melts: Transition from soft spheres to the hard sphere limit}

\author{Sergey A. Khrapak,$^{1,2,3}$ Boris A. Klumov,$^{1,3,4,5}$ and L\'{e}na\"{i}c Cou\"{e}del$^{1}$}

\affiliation{$^1$Aix Marseille University, CNRS, Laboratoire PIIM, Marseille,  France; \\ $^2$Institut f\"ur Materialphysik im Weltraum, Deutsches Zentrum f\"ur Luft- und Raumfahrt (DLR), Oberpfaffenhofen, Germany; \\ $^3$Joint Institute for High Temperatures, Russian Academy of Sciences, Moscow, Russia; \\ $^4$L.D. Landau Institute for Theoretical Physics, Russian Academy of Sciences, Moscow, Russia; \\ $^5$Ural Federal University, Ekaterinburg, Russia}



\begin{abstract}
We study collective modes in a classical system of particles with repulsive inverse-power-law (IPL) interactions in the fluid phase, near the fluid-solid coexistence (IPL melts). The IPL exponent is varied from $n=10$ to $n=100$ to mimic the transition from moderately soft to hard-sphere-like interactions. We compare the longitudinal dispersion relations obtained using molecular dynamic (MD) simulations with those calculated using the quasi-crystalline approximation (QCA) and find that this simple theoretical approach becomes grossly inaccurate for $n\gtrsim 20$. Similarly, conventional expressions for high-frequency (instantaneous) elastic moduli, predicting their divergence as $n$ increases, are meaningless in this regime. Relations of the longitudinal and transverse elastic velocities of the QCA model to the adiabatic sound velocity, measured in MD simulations,  are discussed for the regime where QCA is applicable. Two potentially useful freezing indicators for classical particle systems with steep repulsive interactions are discussed.    
\end{abstract}

\flushbottom
\maketitle
%
%
\thispagestyle{empty}


\section*{Introduction}

The problem of describing collective motion in liquids is one of the long standing major topics in condensed matter physics.~\cite{BoonYip,MarchBook,HansenBook,Copley1975} One of the theoretical approaches to collective motion in neutral liquids is, in certain sense, analogous to that applied to elastic waves in disordered solids. Comparable expressions for the dispersion relations of these elastic waves can be obtained  from the analysis of forth and second frequency moments of the dynamical structure factor $S(k,\omega)$~\cite{HansenBook,deGennes1959} (that is from the sum rules) or from the variational calculation of the frequency spectra of phonons in classical fluids, performed by Zwanzig.~\cite{Zwanzig1967} Particularly enlightening derivation is due to Hubbard and Beeby~\cite{Hubbard1969} who considered their simple theory as either a generalization of the random phase approximation or, alternatively, as a generalization of the phonon theory of solids. Taking this latter viewpoint, the approach was also referred to as the quasi-crystalline approximation (QCA)~\cite{Takeno1971} and we keep this notation for the class of considered theories throughout the paper.  In many cases the QCA-based description is in fair agreement with the dispersion curves deduced experimentally for various simple liquids, at least in the vicinity of the first frequency maximum.~\cite{Hubbard1969,Takeno1971,SingwiPRA1970,KambayashiPRA1992,GiordanoPNAS2010}    

In the context of plasma physics, an analogue of the QCA, known as the quasi-localized charge approximation (QLCA), has been initially proposed as a formalism to describe collective mode dispersion in strongly coupled charged Coulomb liquids.~\cite{GoldenPoP2000} In the last couple of decades the QCA/QLCA approach has been numerously applied to collective wave spectra in one-component-plasma (OCP) in both two-dimensions (2D) and three-dimensions (3D),~\cite{GoldenPoP2000,GoldenPRA1990,GoldenPRA1992,KhrapakOCP2D} magnetized 3D OCP,~\cite{OttPRL2012,OttPRE2013}
2D and 3D Yukawa systems in the context of complex (dusty) plasmas,~\cite{RosenbergPRE1997,KalmanPRL2000,KalmanPRL2004,
DonkoJPCM2008,HouPRE2009,KhrapakPoP2016} 2D dipole systems,~\cite{GoldenPRB2008} etc. In general, the QCA/QLCA model has been documented to describe rather good collective oscillation spectra for soft repulsive interactions between charged particles in the regime of strong coupling (when the potential energy of interaction is much greater than the particle kinetic energy). Interestingly, being developed for the strongly coupled liquid state, QCA reduces to the conventional phonon-dispersion relation in the special case of cold crystalline solid  and can describe real part of the dispersion of the electron (Langmuir) and ion (ion acoustic) waves in classical weakly-coupled (gaseous) plasmas.        

Thus, QCA and (using present notation) its variations have been applied to various real and model systems in a wide interdisciplinary context. It is the appealing simplicity combined with reasonable accuracy, which made QCA a commonly used approximation in studies related to condensed matter, soft matter, and plasma physics. Within the QCA the dispersion relations are explicitly given in terms of the radial distribution function and the pair interaction potential by means of very compact expressions [see Eqs.~(\ref{w_L}) and (\ref{w_T}) below]. More involved theories are available, but they require more information about systems under investigation. For example, the mode coupling theory (MCT)~\cite{Gotze1998} can also successively describe collective excitations of simple classical liquids.~\cite{Gotze1975,Bosse1978_1,Bosse1978_2} However, compared to QCA, it requires additionally information about three-particle correlations and the calculation procedure is much more complex.~\cite{Bosse1978_1} Although MCT demonstrates some improvement over QCA, in particular in the vicinity of the first (roton) minimum of the longitudinal dispersion curve,~\cite{Bosse1978_2} QCA can still be considered as a very useful zero approximation to the dispersion relation.            

Some deficiencies of the QCA approach (in the original Hubbard and Beeby formulation) are well recognized. For example, it does not include direct thermal effects associated with the free movement of the particles in a liquid (although this is not expected to be an important issue at sufficiently strong coupling). Damping effects are also excluded in the original theory (in complex plasmas, for instance, the collisions between charged particles and neutral atoms or molecules represent an important damping mechanism, which has to be accounted for). Possible modifications to QCA in order to remove these deficiencies have been discussed in the literature~\cite{RosenbergPRE1997,HouPRE2009,KhrapakOnset}  and demonstrated to be helpful in certain situations. Overall, it is well documented that the QCA model is a reliable and accurate tool to describe collective modes for strongly interacting particles, provided the interaction potential is soft enough (e.g., in plasma-related systems).

An important issue is, therefore, related to the applicability limits of QCA from the side of steep (hard-sphere-like) interactions. Recently, it has been observed that the sound velocity in a system of particles interacting via the inverse-power-law (IPL) repulsive potential $V(r)\propto r^{-n}$, calculated using the QCA approach, diverges as $\propto\sqrt{n}$ as the potential exponent $n$ increases.~\cite{KhrapakJCP2016} This tendency contradicts the finite value of the sound velocity in a system of hard spheres.~\cite{RosenfeldJPCM1999} Such contradiction is a strong indication that QCA, being a very good approach for soft interactions, loses its accuracy and eventually applicability as the softness of the interaction potential decreases. The purpose of the present paper is to provide an approximate location of the applicability limits of QCA-based approaches. 
  
The most direct way is chosen in this study. 
We consider the IPL family of potentials,
\begin{equation}\label{IPL}
V(r)=\epsilon (\sigma/r)^n,
\end{equation}  
where $\epsilon$ is the energy scale and $\sigma$ is the length scale, and perform extensive molecular dynamics (MD) simulations of the static and dynamical properties of the IPL fluids by increasing the exponent $n$ from $n=10$ to $n=100$. The phase state of the studied IPL systems corresponds to a dense fluid, just at the boundary of the fluid-solid coexistence region (hence we call it melt throughout the paper). One of the output of the simulations is the calculated current fluctuation spectra, which contains information about the dispersion of collective modes. We then compare the dispersion curves from MD simulations with those calculated with the help of QCA approach and identify the condition when significant deviations emerge.  As $n$ increases, the steepness of the potential (\ref{IPL}) increases and in the limit $n\rightarrow \infty$ it approaches the interaction of hard spheres (HS) of diameter $\sigma$. Thus, the chosen strategy seems appropriate for the main purpose of the present study. Note that collective excitations in the IPL fluid with $n=12$ have been recently investigated~\cite{BrykPRE2014} from a somewhat different perspective. 


\section*{Results}\label{Results}

\begin{figure}
\centering
\includegraphics[width=8cm]{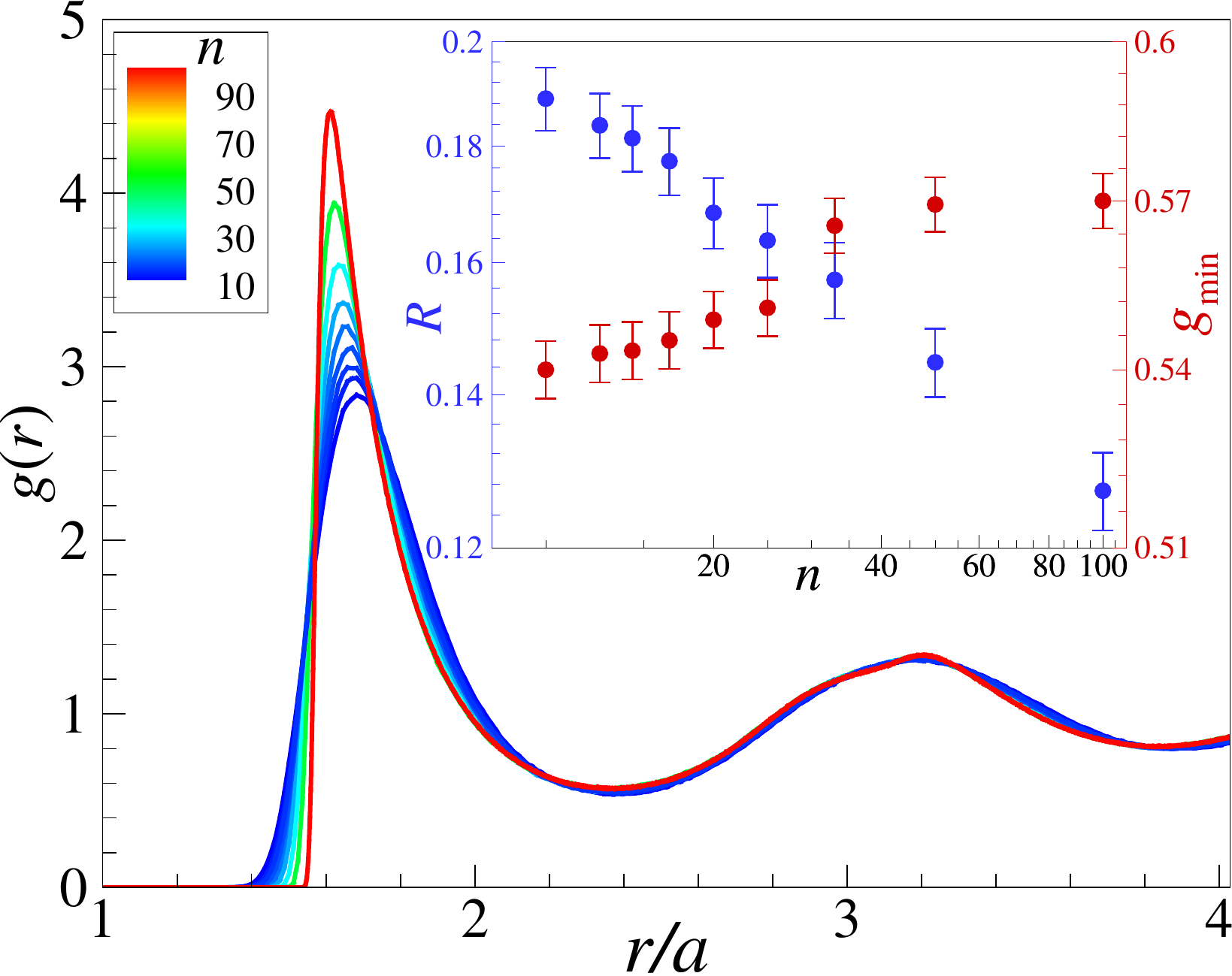}
\caption{Radial distribution functions, $g(r)$, of IPL fluids near the fluid-solid (freezing) phase transition. The curves are color-coded according to the increasing value of the IPL exponent $n$ ($n\sim 10$ correspons to blue, $n\sim 100$ corresponds to red). The inset shows the dependence of the two freezing indicators on the exponent $n$. The blue symbols (left axis) correspond to the familiar Ravech{\'{e}}-Mountain-Streett ratio $g_{\min}/g_{\max}$ at freezing.~\cite{RMS}  The red symbols correspond to the value of the RDF at the first non-zero minimum, $g_{\min}$. We observe that the later quantity is contained in a very narrow range at freezing of the IPL systems. }
\label{Fig1}
\end{figure}

The equilibrium radial distribution function $g(r)$ is required as an input to calculate the dispersion relations of collective modes within the QCA approach (see Methods for details). Thus, we briefly address some important structural properties first. Figure \ref{Fig1} shows the radial distribution functions, $g(r)$, obtained from numerical simulations. The distances are expressed in units of the Wigner-Seitz radius, $a=(4\pi\rho/3)^{-1/3}$, throughout the paper. We observe that the long-range part is very little affected by the variation of the potential steepness. On the other hand, the shape of the first coordination shell [region around the first maximum of $g(r)$] is rather sensitive to the potential steepness. In particular, we see that the value of $g(r)$ at the first maximum increases considerably with $n$. 

The familiar Ravech{\'{e}}-Mountain-Streett freezing rule~\cite{RMS} postulates that the ratio of the first non-zero minimum to the first maximum of $g(r)$, that is $R=g_{\min}/g_{\max}$, remains constant at crystallization. This rule was, however, put forward to describe freezing of the Lennard-Jones fluid and it is not very useful for the IPL family of potential studied here. We see in the inset in Figure~\ref{Fig1} that the ratio $R$ drops by almost a factor of two when $n$ increases from $10$ to $100$. Similar observation was previously reported.~\cite{Agrawal1995} We also see, however, that the value of $g(r)$ at the first minimum  remains practically constant, $g_{\rm min}\simeq 0.55\pm 0.02$. Previously, the value of $g_{\rm min}$ has been used to characterize the fluid-solid (freezing) phase transition in the Lennard-Jones system~\cite{KlumovJPP2013,KlumovJETPl2013} and the glass transition in metallic alloys.~\cite{Ryltsev2016} To which extent this simple freezing indicator applies to other simple model fluids should be investigated in future work.   
 
\begin{figure*}
\centering
\includegraphics[width=17cm]{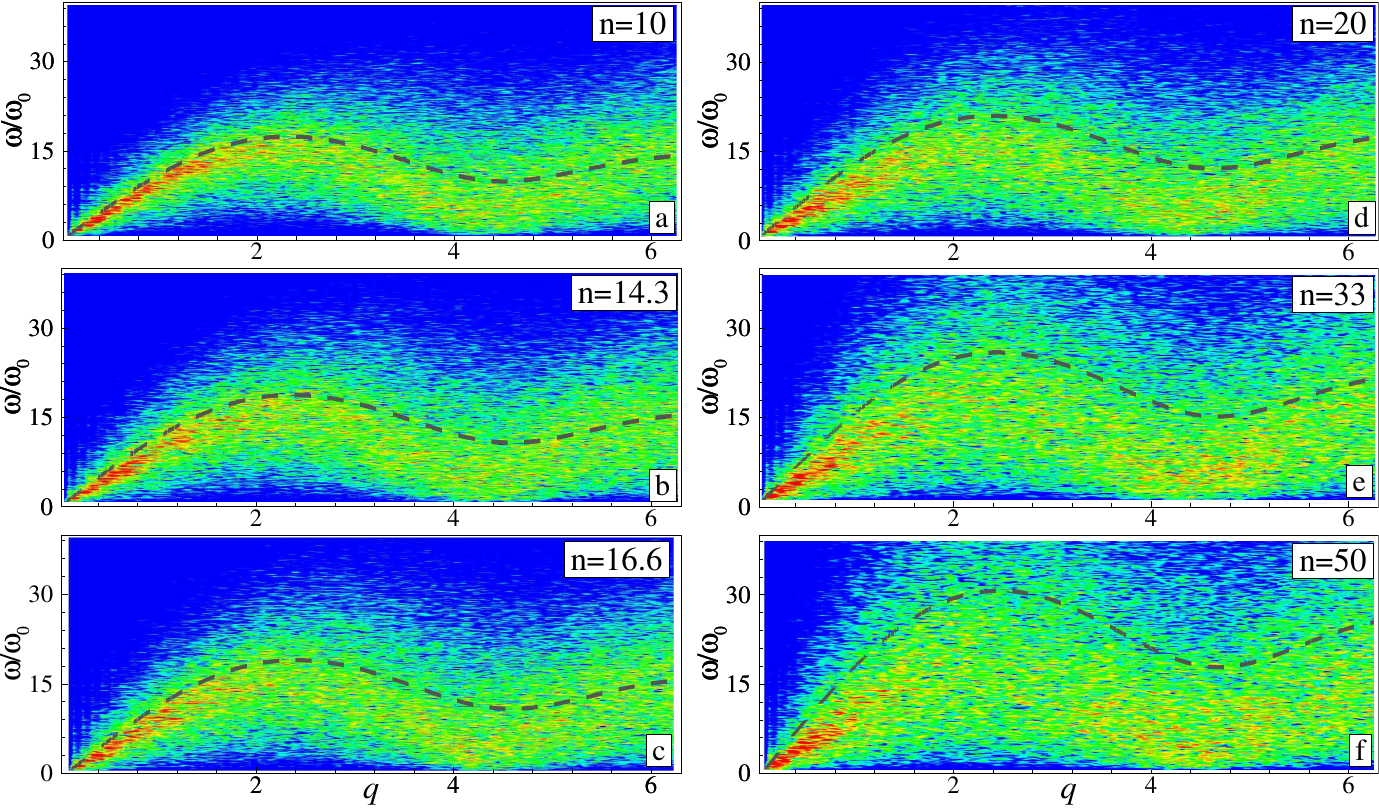}
\caption{Dispersion relations of the longitudinal mode in IPL melts. The frequency is measured in units of $\omega_0=v_{\rm T}/a$, where $v_{\rm T}=\sqrt{T/m}$ is the thermal velocity, the wave-number is in units of $a^{-1}$, $q=ka$. The color background corresponds to the longitudinal current fluctuation spectrum from MD simulations (the intensity increases from blue to red). The dashed lines are the results of calculation using QCA model with $g(r)$ obtained in MD simulations. The results are shown for six different IPL exponents: $n=10$ (a), $n\simeq 14.3$ (b), $n\simeq 16.6$ (c), $n=20$ (d), $n\simeq 33$ (e), and $n=50$ (f).}
\label{Fig2}
\end{figure*}

The dispersion relations of the longitudinal modes in IPL melts are shown in Fig.~\ref{Fig2}. The color background corresponds to the spectral decomposition of the longitudinal current fluctuations. The maximum magnitude (red color) marks the location of dispersion curves as measured in the MD numerical experiment. The black dashed curves correspond to the calculations based on the QCA model with the radial distribution functions taken from MD simulations. Reasonable agreement between QCA results and the current fluctuations analysis is observed in the vicinity of the first frequency maximum for sufficiently soft interactions with $n=10$ (a), $n\simeq 14.3$ (b), and $n\simeq 16.6$ (c). For steeper potentials the deviations are very well observable for $n=20$ (d) and $n\simeq 33$ (e). Finally, for a very steep interaction with $n=50$ (f) the QCA curve is completely off MD simulation results.

\begin{figure}
\centering
\includegraphics[width=8cm]{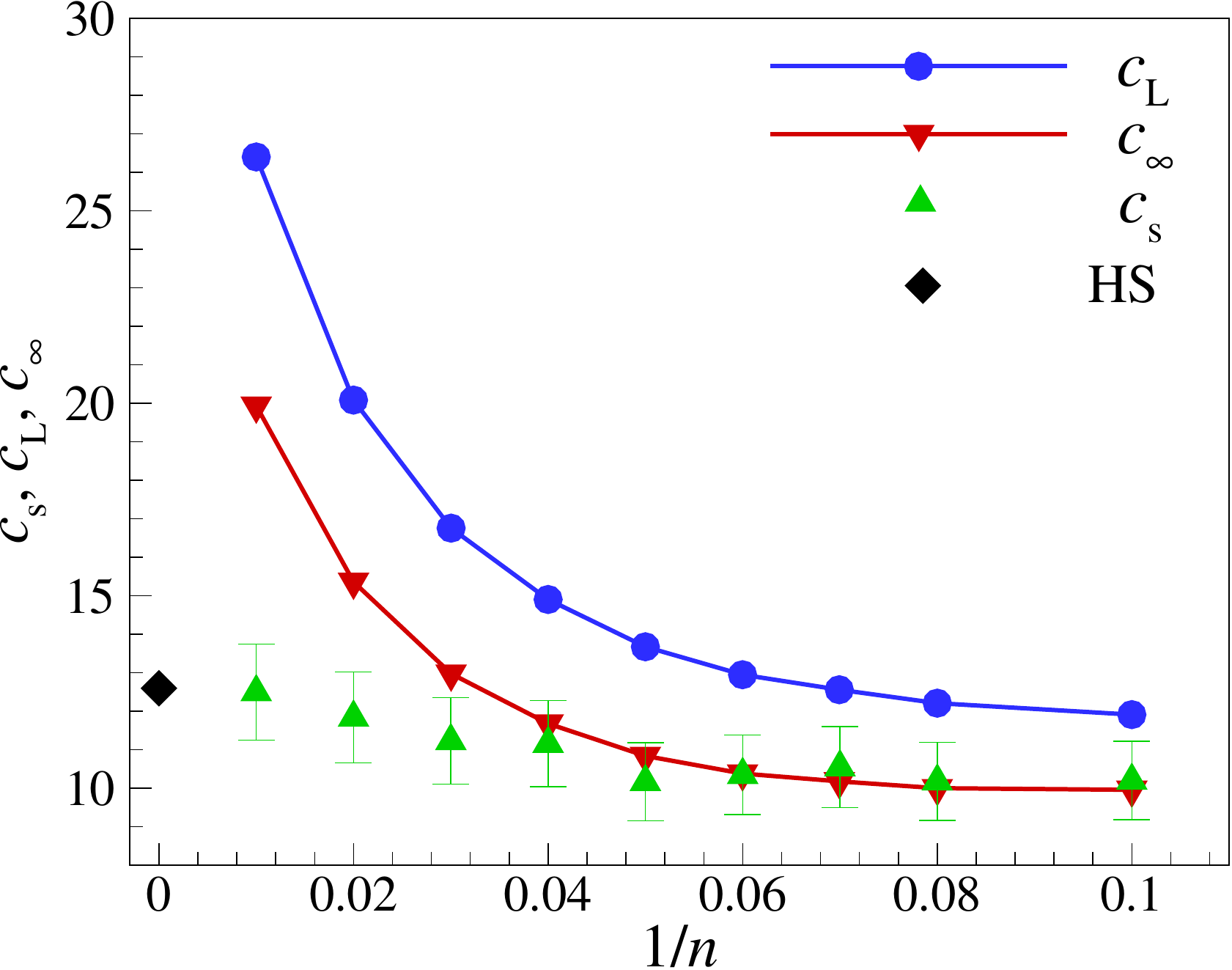}
\caption{Sound velocity of IPL melts versus the inverse IPL exponent $1/n$ (softness parameter). The green data (up triangles) correspond to the (adiabatic) sound velocities obtained directly from MD simulations ($c_{\rm s}$), the blue data (circles) are the sound velocities  obtained using QCA model ($c_{\rm L}$), and the red data (down triangles) show the instantaneous sound velocities $c_{\rm \infty}$. The black diamond at $n=\infty$ corresponds to the adiabatic sound velocity of the hard sphere fluid at the fluid-solid coexistence. All velocities are expressed in units of the thermal velocity $v_{\rm T}$.}
\label{Fig3}
\end{figure}

In Figure~\ref{Fig3} we compare the sound velocities obtained using different approaches. The black diamond, corresponding to $n=\infty$, is the adiabatic sound velocity of the hard sphere system at the fluid side of the fluid-solid coexistence.~\cite{RosenfeldJPCM1999} The green symbols ($c_{\rm s}$) correspond to the linear fit of the MD data on current fluctuation spectra in the long-wavelength limit ($q\rightarrow 0$). The blue data ($c_{\rm L}$) correspond to the extrapolation of the QCA dispersion relation in the long-wavelength limit. The red data ($c_{\infty}$) show the related instantaneous sound velocity, $c_{\infty}=c_{\rm L}\sqrt{\tfrac{5(3+n)}{3(1+3n)}}$. All the sound velocities are expressed in units of the thermal velocity $v_{\rm T}$. It is observed that $c_{\rm s}$ and $c_{\infty}$ practically coincide for $n\lesssim 25$, while $c_{\rm L}$ overestimates these sound velocities by approximately a factor of $3/\sqrt{5}$ in the considered regime. These observations are discussed in detail in the next Section.  

\section*{Discussion}\label{Discussion}

One of the main observations is that the QCA model, designed to describe elastic modes in simple fluids, has limited applicability with respect to the interactions softness. It is only applicable for sufficiently soft interactions. In terms of the IPL family studied here, the potential exponent $n$ should satisfy $n\lesssim 20$.

Even when the QCA reasonably describes the dispersion near the first frequency maximum, the predicted elastic sound velocity $C_{\rm L}$ is generally higher than the adiabatic sound velocity $C_{\rm s}$ (see Fig.~\ref{Fig3}). The magnitude of $C_{\rm s}$ can still be estimated using the QCA by means of the instantaneous sound velocity $C_{\infty}^2=C_{\rm L}^2-\tfrac{4}{3}C_{\rm T}^2$, related to the instantaneous bulk modulus (see section Methods for details). For sufficiently large exponent $n$, the ratio of elastic to instantaneous sound velocities is approximately $3/\sqrt{5}$. When the potential becomes softer this ratio decreases and reach unity for $n=3$. The case $n=3$ is special for the IPL model since for $n\leq 3$ a uniform neutralizing background should be applied.~\cite{DubinPRB1994} In addition, the dispersion relations for longitudinal waves are non-acoustic in the long-wavelength limit for $n=1$ and $n=2$.~\cite{GoldenPoP2000,Fingerprints}  
A general remark regarding the regime of soft interactions is appropriate here. In soft interacting dense fluids, not too far from the fluid-solid transition, the longitudinal sound velocity is normally much higher than the transverse sound velocity, $C_{\rm L}\gg C_{\rm T}$. This implies that $C_{\infty}\simeq C_{\rm L}$ and QCA provides a rather good estimate of the actual thermodynamic sound velocity. The very fact that the QCA elastic and thermodynamic sound velocities are rather close to each other has been numerously documented for weakly screened Yukawa (screened Coulomb) systems at strong coupling.~\cite{QCA_Relations,KhrapakPRE03_2015,KhrapakPPCF2016}   

It is observed in Fig.~\ref{Fig3} that there is a very wide range of potential softness, at least from $n=10$ to $n=\infty$, where the longitudinal sound velocity near freezing varies weakly when approaching its HS asymptotic value ($\simeq 12.5$). This is unlikely to be a specific property of the IPL system. For example, it is well recognized that the ratio of the sound to thermal velocity of many liquid metals and metalloids has about the same value $\simeq 10$ at the melting temperature.~\cite{IidaBook,BlairsPCL_2007} This observation has been already discussed in the context of the HS system~\cite{RosenfeldJPCM1999} and, more recently, in the context of soft-sphere interactions.~\cite{KhrapakJCP2016} We just remark here that the approximate constancy of the sound velocity (in units of thermal velocity) near the fluid-solid phase transition can perhaps be of some use as an approximate dynamical freezing indicator for systems with steep repulsive interactions.  

\begin{figure}
\centering
\includegraphics[width=8cm]{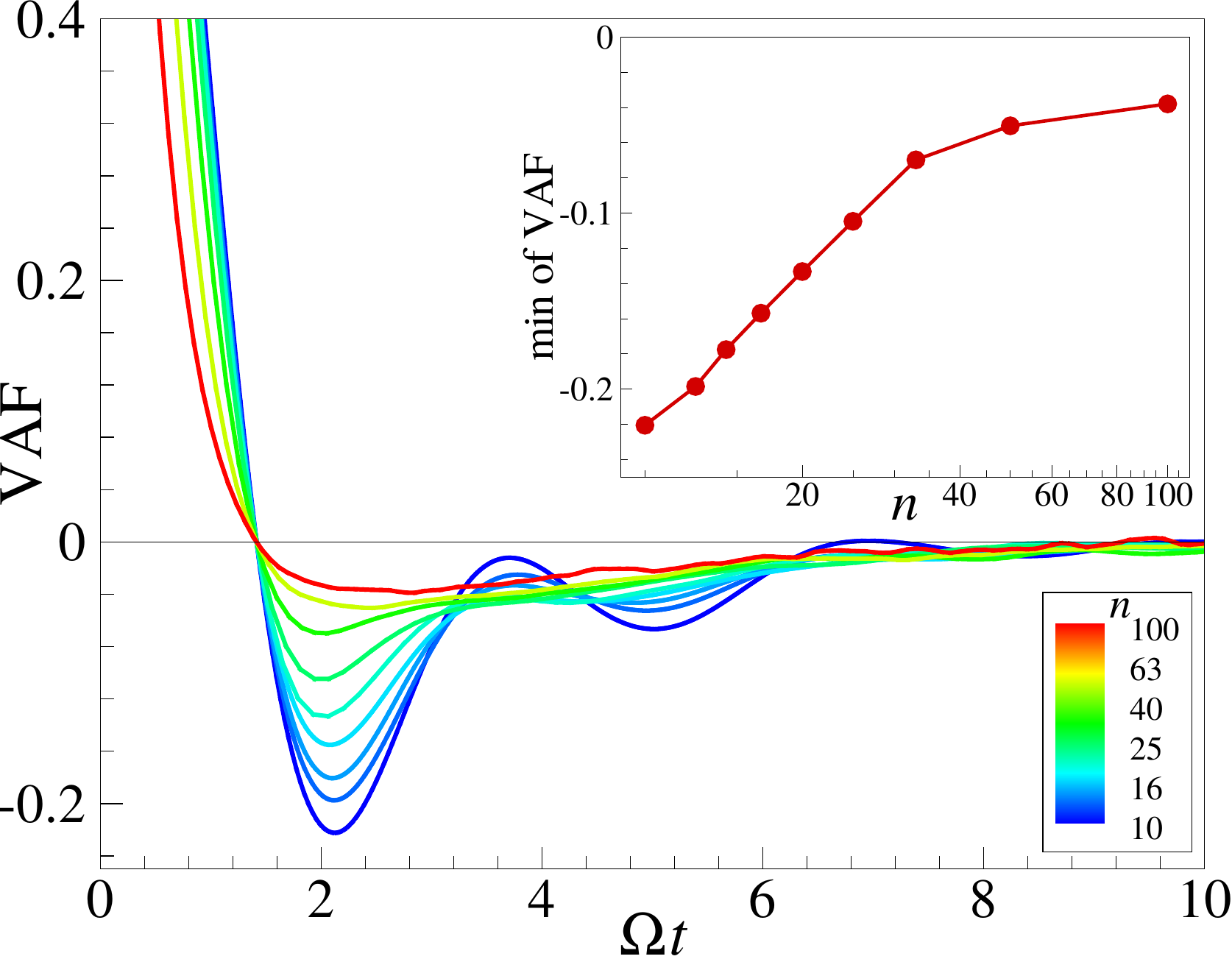}
\caption{Velocity autocorrelation functions (VAF) of the IPL melts when increasing the exponent $n$ from $n\simeq 10$ to $n\simeq 100$ (curves are color-coded correspondingly). The time is normalized in such a way, that all VAF curves intersect zero at the same point, $\Omega$ is roughly the Einstein frequency.~\cite{HansenBook} The inset shows the dependence of the minimum value of VAF on the IPL exponent $n$, revealing the saturation-like behavior at large $n$.}
\label{Fig4}
\end{figure}

\begin{figure*}
\centering
\includegraphics[width=12cm]{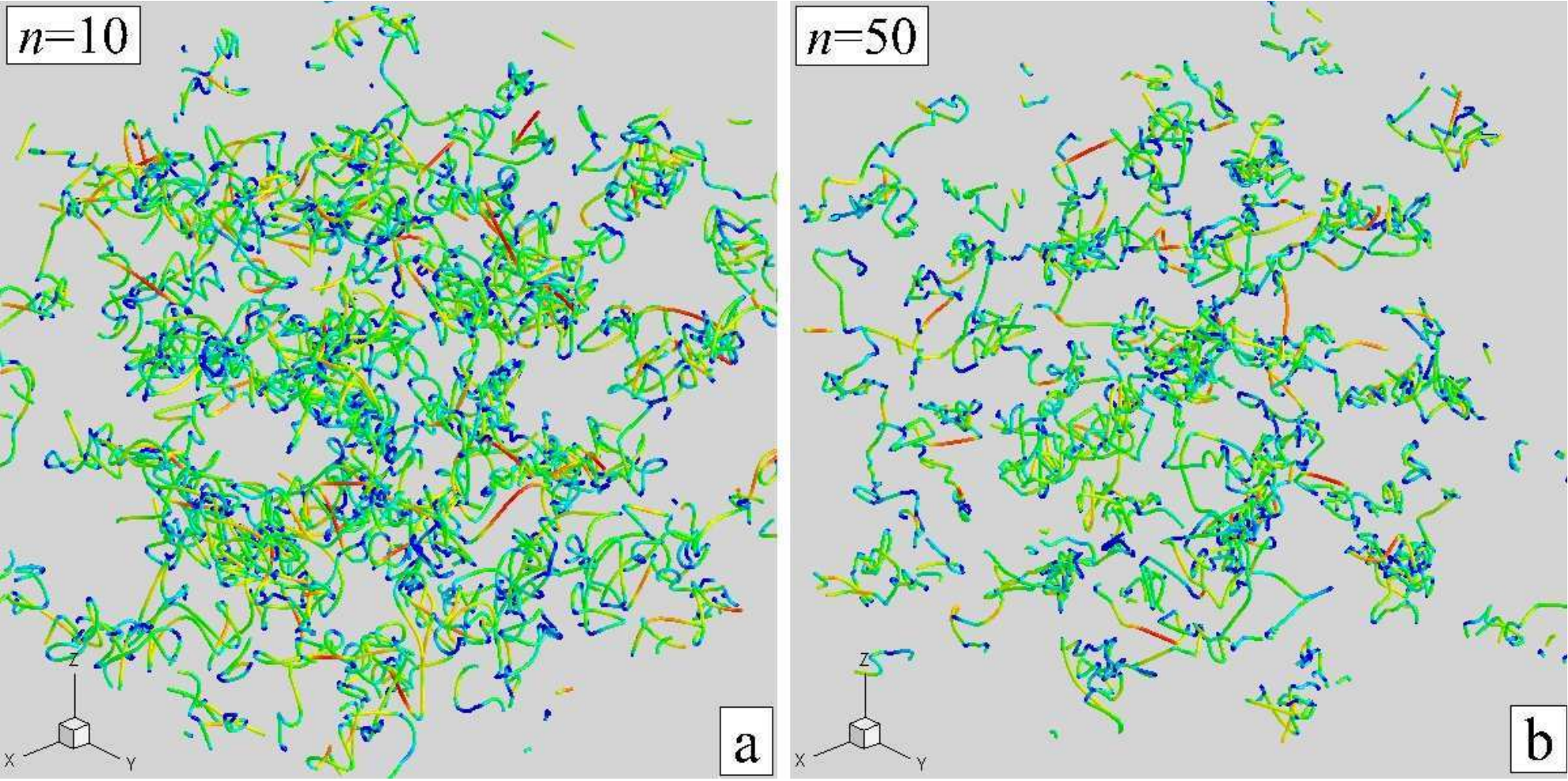}
\caption{Particle trajectories within a cubic box with the side length $\simeq 5a$, chosen near the center of the simulation cell, demonstrating the particle motion in IPL melts for $n=10$ (a) and $n=50$ (b). The trajectories are color-coded according to particle energy (transition from red to blue corresponds to the decrease in energy). The time interval during which the trajectories are followed corresponds roughly to one hundred of inverse Einstein frequencies.}
\label{Fig5}
\end{figure*}

\begin{figure}
\centering
\includegraphics[width=8cm]{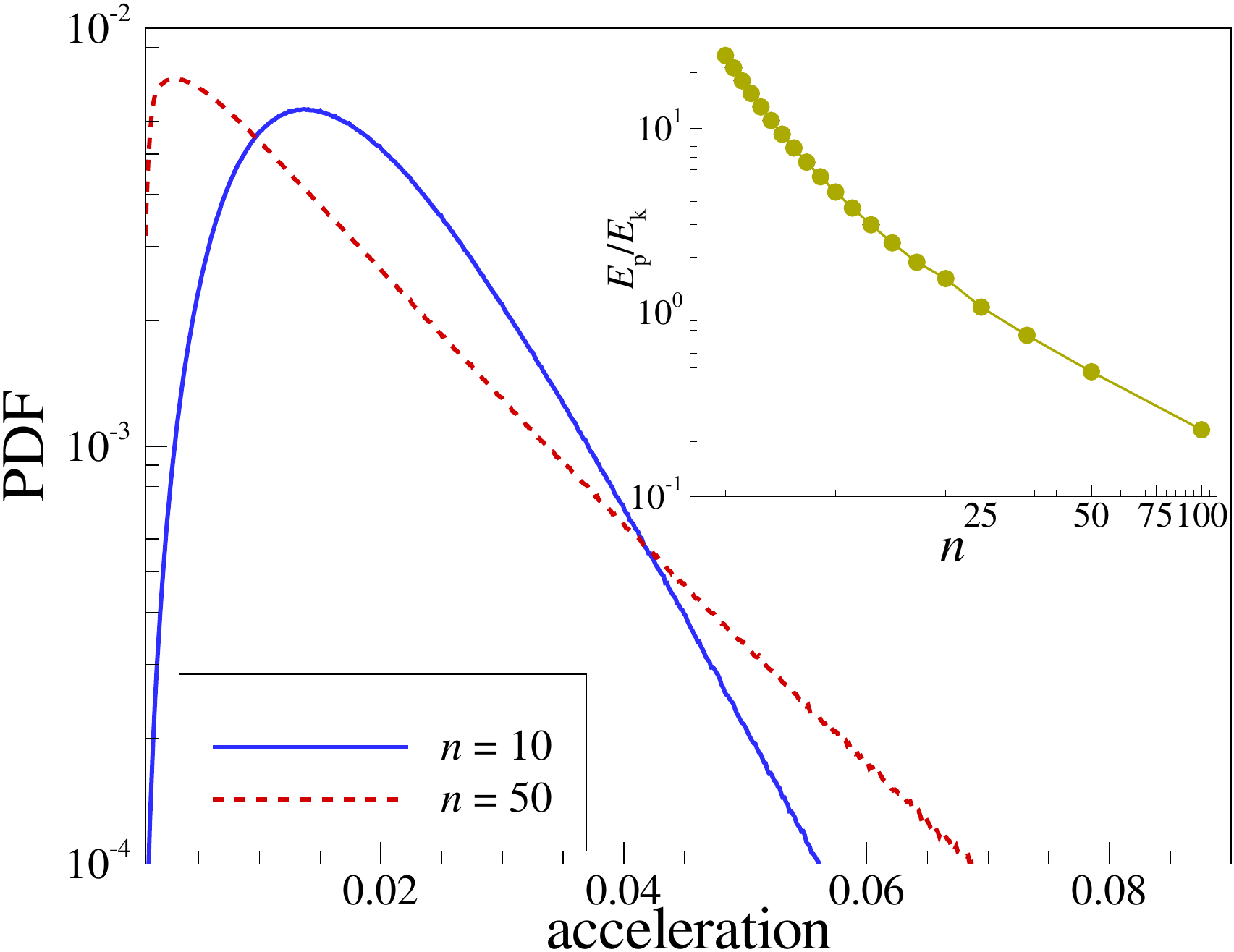}
\caption{Probability distribution functions (PDF) of the particle accelerations in IPL melts for two different exponents: $n=10$ (blue solid curve) and $n=50$ (red dashed curve). 
The inset shows the ratio of the potential to kinetic energies stored in the system of interacting particles. The ratio drops below unity at $n\simeq 30$, which is close to the estimated failure of the QCA. For $n > 30$ the IPL melts are strongly correlated, but not strongly coupled.  
}
\label{Fig6}
\end{figure}

In an attempt to understand the failure of the QCA consideration in the limit of steep interactions we evaluated some additional dynamical properties of the system. The results are summarized in Figs.~\ref{Fig4} -- \ref{Fig6}.

Figure \ref{Fig4} demonstrates how the velocity autocorrelation functions (VAF) vary upon increasing the potential steepness. The shape of VAFs is changed considerably as $n$ increases. For $n\sim 10$ a clear ``caging'' behaviour of particles trapped and oscillating in slowly fluctuating potential wells, reminiscent particle dynamics in OCP,~\cite{DonkoPRL2002} is observed. For $n\sim 100$ these oscillations are heavily suppressed and the shape of the VAF tends to that of the HS fluids near the freezing point.~\cite{WilliamsPRL2006} The amplitude of the first minimum of the VAF shifts considerably upwards with the increase of $n$ (see inset). Thus, a pronounced oscillatory caging behaviour seems important for the application of the QCA, in qualitative agreement with the physical picture behind the development of the QLCA model.~\cite{GoldenPoP2000} 

Related to this is the observation by Brazhkin {\it et al.}~\cite{BrazhkinPRE2012,BrazhkinUFN2012} that the caging dynamics changes significantly upon increasing the potential steepness of the IPL potential. Namely, they calculated the potential-to-kinetic energy ratio of the IPL fluids near the melting curve and found that it drops below unity at $n\sim 30$. We repeated this calculation using the data summarized by Agrawal and Kofke~\cite{Agrawal1995} and present the obtained results in the inset of Figure~\ref{Fig6}. Based on the observations, the particle motion inside the cages can be characterized as harmonic in the soft interaction regime and purely collisional absolutely non-harmonic in the steep interaction regime.~\cite{BrazhkinPRE2012} The transition between these two regimes at $n\sim 30$ is close to the point where QCA failure occurs. The dynamical crossover is likely to be the reason.  

In Figure \ref{Fig5} individual particle trajectories in a small cubic box, placed near the center of the computation cell, are shown for two cases, $n=10$ and $n=50$. The trajectories are color coded according to  the particles energy. Trajectories shown in (a) appear relatively smooth and continuous. In contrast, we observe in (b) strong breaks in particles trajectories for steeper potential with $n=50$, which result from short hard-core-like collision events. This is further illustrated in Fig.~\ref{Fig6}, which shows probability distribution functions for accelerations (in arbitrary units). There is a significant tail corresponding to higher accelerations in the case $n=50$, compared to the case $n=10$. This implies, quite expectedly, that as the interaction steepens, the number of short time collisions with higher accelerations increases and hence the characteristic collision time decreases. In this context, it should be pointed out that the necessity of  a careful consideration of the time-scale of two-particle collisions was discussed already by Schofield.~\cite{Schofield1966} In particular, he argued that if there is a sharp hard core, then the duration of a collision $\tau_{\rm D}$ tends to zero. If high frequency elastic solid-like excitations exist, they can only appear above a background of width $\tau_{\rm D}^{-1}$.~\cite{Schofield1966}

One more consideration can be based on the observed behaviour of the instantaneous bulk modulus. Figure~\ref{Fig3} demonstrates that the actual adiabatic bulk modulus, as inferred from the sound velocities measured in MD simulations, remains finite and smoothly approaches its HS limiting value. In contrast, the high frequency bulk modulus diverges as $K_{\infty}\propto \sqrt{n}$ when $n$ increases. It is instructive to consider the main steps of the derivation of the analytical expression for $K_{\infty}$.~\cite{Zwanzig1965,Schofield1966} A particularly simple route to derive its excess (associated with potential interactions between the particles) part has been discussed previously.~\cite{SingwiPRA1970,KhrapakOnset} The starting point is the virial expression for the excess pressure, which should be differentiated with respect to the particle density. In doing so a term containing an explicit derivative $\partial g(r)/\partial \rho$ appears. The latter clearly depends on the thermodynamic process. In the infinite frequency (instantaneous) limit one assumes that no relaxation is allowed and the density change occurs without any structural rearrangement. Mathematically, this implies that $g(r)$ does not change if $r$ is scaled by $\rho^{-1/3}$. This allows to express $\partial g(r)/\partial \rho$ in terms of $\partial g(r)/\partial r$ and then, integrating by parts, to recover the conventional results.~\cite{Zwanzig1965,Schofield1966} It is the assumption of no structural rearrangement [independence of $g(r\rho^{1/3})$ of $\rho$] that causes problems. While well justified for sufficiently soft interactions, it is clearly not applicable to HS like interaction, because an intrinsic length scale -- the hard sphere diameter -- emerges. Thus, it is legitimate to say that the divergence of the high frequency elastic moduli in the limit of very steep interactions is unphysical. Rather, the conventional expressions must not be applied in this limit.      

Of course, all these considerations are merely qualitative. Only direct comparison between the dispersion relations from MD simulations and the QCA model allowed us to quantify the limits of the applicability of elastic approaches to collective modes in fluids as $n\lesssim 20$ for the IPL model. On the other hand, none of these considerations is directly related to the special shape of the IPL potential studied here. Therefore, one should expect that elastic approaches such as QCA also fail for other systems of strongly interacting particles, provided the interaction potential becomes too steep.

The main results of our investigation can be formulated as follows:

(i) The QCA description of collective modes fails for sufficiently steep interactions; For the IPL model considered here this failure occurs at $n\sim 20$.

(ii) For softer potentials QCA can provide a reasonable description of the high-frequency portion of the longitudinal dispersion relation (in particular, in the vicinity of the first maximum); However, the QCA elastic sound velocity overestimates the actual sound velocity. 

(iii) The instantaneous sound velocity, evaluated using the instantaneous bulk modulus, is an appropriate measure of the actual (adiabatic) sound velocity observed in MD simulations; This instantaneous sound velocity ($C_{\infty}$) is related to the elastic longitudinal ($C_{\rm L}$) and transverse ($C_{\rm T}$) sound velocities of the QCA approach via $C_{\infty}^2=C_{\rm L}^2-\tfrac{4}{3}C_{\rm T}^2$.

(iv) In the limit of soft interactions (not considered here, but previously extensively studied  using Coulomb and Yukawa model systems), the strong inequality $C_{\rm L}\gg C_{\rm T}$ generally holds, which implies that the longitudinal QCA elastic sound velocity, instantaneous sound velocity, and the thermodynamic sound velocity are all close to each other;

(v) The conventional expressions for high frequency (instantaneous) elastic moduli should not be employed in the hard-sphere interaction limit, where they diverge.

The main finding can also be briefly reformulated as follows. The QCA model was formulated originally to describe collective motion in simple fluids without any explicit limitations regarding the shape of the interaction potential. It turns out, however, that this model works very well for soft interactions (in particularly in the plasma-related context), but fails when the interaction becomes sufficiently steep. We located this failure here using the IPL family of potentials.   

In addition to this, two important relevant observation have been reported.             
It was found that the value of radial distribution function $g(r)$ at the first minimum remains practically constant for all investigated IPL fluids with $100>n>10$ near the fluid-solid coexistence. This quantity is shown to exhibit much smaller relative variation than the ratio of the first minimum to the first maximum of $g(r)$ appearing in the Ravech{\'{e}}-Mountain-Streett freezing criterion. For the same range of steepness, the ratio of the sound to thermal velocities also remains practically constant, slowly increasing towards the hard-sphere limiting value as $n$ increases. These two observations can potentially be used as approximate structural and dynamical freezing indicators for simple systems with steep repulsive interactions. 

\section*{Methods}

{\bf QCA dispersion relations.} Within the QCA approach the dispersion relations of the longitudinal and transverse modes are written as~\cite{Hubbard1969,Takeno1971} 
\begin{equation}\label{w_L}
\omega_{\rm L}^2=\frac{\rho}{m}\int\frac{\partial^2 V(r)}{\partial z^2} g(r) \left[1-\cos(kz)\right]d{\bf r},
\end{equation} 
\begin{equation}\label{w_T}
\omega_{\rm T}^2=\frac{\rho}{m}\int\frac{\partial^2 V(r)}{\partial x^2} g(r) \left[1-\cos(kz)\right]d{\bf r},
\end{equation}
where $\rho$ is the particle number density, $m$ is the particle mass, $\omega$ is the frequency, $k$ is the wave number, and $z=r\cos\theta$ is the direction of the propagation of the longitudinal wave. The frequency spectra are thus determined by the first and second derivatives of the interaction potential $V(r)$ and the equilibrium radial distribution function of the fluid $g(r)$; the explicit expression for $\omega_{\rm L}$ (after intergration over angles) can be found in the literature.~\cite{Fingerprints,Nossal1968} 

As displayed in Eqs.~(\ref{w_L}) and (\ref{w_T}), the dispersion relations within QCA are completely determined by the potential interaction between the particles. In related approaches (e.g. based on sum rules or generalized high-frequency elastic moduli) kinetic terms can naturally appear additively to the potential terms ($3k^2v_T^2$ for the longitudinal mode and $k^2v_T^2$ for the transverse mode, where $v_{\rm T}=\sqrt{T/m}$ is the thermal velocity).~\cite{HansenBook,Zwanzig1967} Shofield argued that these kinetic terms should not be included in the dispersion relation.~\cite{Schofield1966} We can avoid further related discussion here by pointing out that the kinetic contributions are normally numerically small at liquid densities.~\cite{Schofield1966} Since in the present paper we consider very dense liquids, just in the vicinity of the solidification, it would be more than sufficient to keep only the potential terms in the dispersion relations ~(\ref{w_L}) and (\ref{w_T}).  

In the limit of long wavelengths the dispersion relations of the IPL fluid exhibit an acoustic dispersion and the longitudinal and transverse sound velocities can be introduced,
\begin{equation}
\lim_{k\rightarrow 0} \frac{\omega_{\rm L/T}^2}{k^2} = C_{\rm L/T}^2.
\end{equation}      
Further, for the IPL systems these sound velocities can be easily related to the reduced excess pressure of the fluid, $p_{\rm ex}=P/\rho T -1$, as follows:
\begin{equation}
C_{\rm L}^2=(3n+1)v_{\rm T}^2 p_{\rm ex}/5,
\end{equation} 
and
\begin{equation}
C_{\rm T}^2=(n-3)v_{\rm T}^2 p_{\rm ex}/5.
\end{equation}
Note the relationship $C_{\rm L}^2-3C_{\rm T}^2=2p_{\rm ex}v_{\rm T}^2$, which is a general (independent of the interaction potential) property of the QCA model in both two- and three dimensions.~\cite{QCA_Relations} In addition, the longitudinal and transverse sound velocities are related to the high frequency shear modulus $G_{\infty}$ and the high frequency bulk modulus $K_{\infty}$,~\cite{Zwanzig1967,Zwanzig1965}
\begin{equation}
C_{\rm L}^2=(\tfrac{4}{3}G_{\infty}+K_{\infty})/m\rho, \quad\quad
C_{\rm T}^2=G_{\infty}/m\rho.
\end{equation}
This is very similar to the conventional relations for elastic waves in an isotropic medium.~\cite{LL_Elasticity,Trachenko2015} 
The ``instantaneous'' sound velocity can be introduced using the high frequency (instantaneous) bulk modulus~\cite{Schofield1966}
\begin{equation}
C_{\infty}^2=K_{\infty}/m\rho=C_{\rm L}^2-\tfrac{4}{3}C_{\rm T}^2.
\end{equation} 
This instantaneous sound velocity $C_{\infty}$ appears to be rather close to the conventional thermodynamic sound velocity for soft repulsive potentials, as has been documented for instance for the case of strongly coupled Yukawa fluids.~\cite{QCA_Relations} But this is not the case for steeper interactions as demonstrated in this paper.  

{\bf Numerical simulations.} Simulations were performed on graphics processing unit (NVIDIA Tesla K80) using the HOOMD-blue software.~\cite{HOOMD,Anderson2008} We used $N=55296$ particles in a cubic box with periodic boundary conditions. Simulations were performed in the canonical ensemble ($NVT$) using the Langevin thermostat at a temperature $T=\epsilon=1$. A cut-off radius for the potential $r_{\rm max}$ was set such that $V(r_{\rm max})=5\times 10^{-8}$. The numerical time step was set to be approximately $10^{-2}$ of the inverse Einstein frequency for each exponent $n$ investigated.  
The system was first equilibrated for fifteen million time steps,
and then,the particle positions and velocities were saved for 80 000 time steps. 

The IPL exponent varied from $n=10$ to $n=100$. 
For each chosen exponent $n$ the system state corresponded to the fluid side of the fluid-solid (fcc lattice) coexistence region. This was achieved by
choosing the reduced density $\rho\sigma^3$ according to the  data tabulated in the Table 1 of Agrawal and Kofke.~\cite{Agrawal1995}  

For each simulation runs the following diagnostic tools were implemented. We calculated the radial distribution function $g(r)$ and the normalized velocity autocorrelation function (VAF)
\begin{equation}
Z(t)=\frac{\langle{\bf v}(t){\bf v}(0)\rangle}{\langle{\bf v}(0)^2\rangle},  
\end{equation}
where ${\bf v}(t)$ is the velocity of a particle at time $t$ and $\langle\cdot\cdot\cdot \rangle$ denotes an average over all particles. In addition, the longitudinal component of the particle current
\begin{equation}
J_{\rm L}(k,t)=(1/N k) \sum_{i=1}^{N}{\bf k}\cdot{\bf v}_i(t)e^{{\bf k}\cdot{\bf r}_i(t)}
\end{equation}
was calculated. The Fourier transform in time was performed to obtain the current fluctuation spectrum and, hence, the longitudinal dispersion, similarly to our previous approach.~\cite{KhrapakPoP2016}

\bibliography{References_IPL_1}

\section*{Acknowledgements}

This work was supported by the A*MIDEX project (Nr.~ANR-11-IDEX-0001-02) funded by the French Government ``Investissements d'Avenir'' program managed by the French National Research Agency (ANR). Simulations were performed using the granted access to the HPC resources of Aix-Marseille Universit\'e financed by the project Equip@Meso (ANR-10-EQPX-29-01) of the program  ``Investissements d'Avenir'' program managed by the French National Research Agency (ANR). Structure analysis was supported by Russian Science Foundation
(grant RSF 14-50-00124).

\section*{Author contributions statement}

S.K. conceived the research and wrote the manuscript. L. C. conducted simulations. B. K. and L. C. performed structural and dynamical analysis. All authors reviewed the manuscript. 

\section*{Additional information}

\textbf{Competing financial interests:} The authors declare no competing financial interests.

\end{document}